\documentclass{mem}
\usepackage{natbib}
\usepackage{txfonts}
\usepackage{balance}
\usepackage{graphicx}
\usepackage{flushend}
\usepackage[breaklinks,pdftex]{hyperref}
\usepackage{xcolor}

\newcommand{\herakoi}{\texttt{herakoi}}

\begin{document}

\title{
\texttt{herakoi}: a sonification experiment for astronomical data
}


\author{
Michele Ginolfi \inst{1,2}, 
Luca Di Mascolo\inst{3},
Anita Zanella\inst{4}
          }

\institute{
        Dipartimento di Fisica e Astronomia, Università di Firenze, Via G. Sansone 1, I-50019, Sesto Fiorentino (Firenze), Italy
        \and
        INAF — Osservatorio Astrofisico di Arcetri, Largo E. Fermi 5, I-50125, Florence, Italy
\and
Laboratoire Lagrange, Université Côte d'Azur, Observatoire de la Côte d'Azur, CNRS, Blvd de l'Observatoire, CS 34229, 06304 Nice cedex 4, France
\and
INAF, Vicolo dell’Osservatorio 5, I-35122 Padova, Italy\\
}

\authorrunning{Ginolfi, Di Mascolo, Zanella}

\titlerunning{\texttt{herakoi}: a sonification experiment for astronomical data}

\abstract{Recent research is revealing data-sonification as a promising complementary approach to vision, benefiting both data perception and interpretation. We present \herakoi, a novel open-source software that uses machine learning to allow real-time image sonification, with a focus on astronomical data. By tracking hand movements via a webcam and mapping them to image coordinates, \herakoi\ translates visual properties into sound, enabling users to "hear" images. Its swift responsiveness allows users to access information in astronomical images with short training, demonstrating high reliability and effectiveness. The software has shown promise in educational and outreach settings, making complex astronomical concepts more engaging and accessible to diverse audiences, including blind and visually impaired individuals. We also discuss future developments, such as the integration of large language and vision models to create a more interactive experience in interpreting astronomical data.
}
\maketitle{}

\section{Introduction}

Astronomy has traditionally been a visually dominated science, relying heavily on images, graphs, and visual data representations to interpret and communicate observations 
and, in general, scientific findings. However, human vision is inherently limited, sensitive only to a narrow band of the electromagnetic spectrum (from approximately 400 to 700 nanometers) leaving a vast span in wavelengths –– from gamma ray to ultraviolet and from infrared to radio –– beyond our direct perception. To compensate for this limitation, astronomers often transform data from non-visible wavelengths into visual formats using techniques like color mapping. 
While effective to some extent, these methods can lead to information loss and may not fully capture the richness of the original data (\citealp{Harrison2022, Zanella2022}).
Moreover, certain astronomical observations, such as those from radio telescopes 
produce data in the form of Fourier-space information (generally referred to as ``visibilities'') rather than direct images. Converting these complex datasets into images can result in further loss of information and may obscure subtle features critical for scientific analysis (e.g., \citealp{ThompsonRadio}). 
Additionally, much of the Universe consists of phenomena that do not interact with light at all, such as dark matter and dark energy. These components cannot be observed through traditional visual means, making it insufficient to rely solely on sight for a comprehensive understanding of the cosmos (\citealp{Peebles2003, Planck2020}).

In response to these challenges, the field of data sonification has emerged as a promising avenue for enhancing data interpretation and accessibility in astronomy (\citealp{Cooke_Diaz2017}). Sonification involves converting data into sound, leveraging the human auditory system's sensitivity to temporal and frequency patterns to reveal insights that might be less apparent visually. This approach not only complements traditional visual analyses 
for all, but also makes astronomical data more accessible to individuals who are blind or visually impaired (BVI), promoting inclusivity in research, education and public outreach (\citealp{james2022, Harrison2023}).
The importance of sonification in astronomy is multifaceted. It has the potential to improve engagement by providing a novel and immersive way to experience astronomical phenomena. In educational contexts, sonification can aid in teaching complex concepts by appealing to different learning styles and sensory modalities (\citealp{Guiotto2023, Guiotto2024a}). For research, auditory data representation can reveal patterns and correlations not easily discernible through visual inspection alone (\citealp{james2023}). 
Despite these benefits, the field faces limitations, such as the lack of standardization in sonification techniques and the need for tools that are both effective and user-friendly (see e.g., \citealp{Zanella2022}).

Recent advancements in artificial intelligence (AI) and machine learning (ML) have the potential to drive substantial progress in data sonification. These technologies enable more sophisticated and real-time data processing, opening new possibilities for interactive and dynamic sonification applications. 
Leveraging these 
innovations, we have developed \texttt{herakoi}
\footnote{\url{https://github.com/herakoi/herakoi}}, \citep{herakoi} a sonification tool designed for a sound-centric interpretation of visual elements.

\texttt{herakoi} utilizes a ML-based algorithm for real-time hand recognition to track the position of the user's hands via a standard webcam. The coordinates of the detected hand landmarks are re-projected onto the pixel coordinates of a selected astronomical image. As users move their hands over the image, the tool captures the visual properties of the ``touched'' pixels—such as color and saturation—and converts them into corresponding sound properties using any virtual or physical MIDI player of the user's choice. This dynamic interaction allows users to ``hear'' the image, transforming the experience of data exploration into a multi-sensory activity.
This technology not only facilitates more inclusive access to visually driven information but also offers novel pathways for human-machine interaction. 
By engaging multiple senses, \texttt{herakoi} enhances the user's ability to perceive and interpret complex data. Preliminary applications of \texttt{herakoi} have demonstrated significant potential 
(see a discussion in Sec. \ref{sec:discussion}). In educational settings, it has proven to be an effective tool for helping young students grasp astronomical concepts \citep{Guiotto2023, Guiotto2024a}, catering to diverse learning styles. Additionally, \texttt{herakoi} has successfully engaged audiences during public outreach events, serving as a valuable resource for both sighted and BVI individuals.

In this paper we present \texttt{herakoi}, highlighting its capabilities and contributions to the field of astronomical sonification. 
We discuss the underlying AI methods that enable its real-time interaction features and explore its impact on education, outreach, and research. 
\texttt{herakoi} represents a step forward in making astronomy more accessible and engaging for all.

\vspace{-0.3cm}
\section{\texttt{herakoi} -- how does it work?}
\label{sec:herakoialgorithm}

\texttt{herakoi} is a motion-sensing sonification tool that transforms images into sound in real-time. 
It achieves this by employing the publicly available MediaPipe (\citealp{mediapipe2019}) Hand Landmarker model to detect and track 21 key landmarks
\footnote{\url{https://ai.google.dev/edge/mediapipe/solutions/guide}}. 
The model was trained on approximately 30~000 real-world images of hands, supplemented with several synthetic  models superimposed on various backgrounds. The model bundle integrates a palm detection algorithm with the hand landmark detection model, resulting in a nested detection approach that progressively refines the estimation of key coordinates.
Initially, the palm detection algorithm identifies hands within any input image, defining a region of interest. Within this region, the hand landmark detection model pinpoints specific hand landmarks. The MediaPipe Hand Landmarker can operate on both static images and continuous data streams. 
In our application, we use it to track the user's hand positions in real-time via a webcam connected to the computer.
The model captures the coordinates of the user's hands, which are then mapped onto the pixel coordinates of a selected image.
\texttt{herakoi} offers two options for defining the ``touched'' area, catering to different user needs.
The first option defines a square region centered around the user's index finger, with a customizable side length specified in pixels. This allows for precise and controlled interaction, suitable for applications requiring exactness. The second option defines a rectangular region based on the coordinates of index finger and thumb. As the distance between the two 
changes, the size and aspect ratio of the rectangle adjust accordingly. This enables dynamic and intuitive interaction with the image, allowing users to naturally zoom in and out by varying the spacing between their fingers.

Once the ``touched'' pixel area is defined, its visual properties are converted into sound properties of a selected instrument from a virtual MIDI keyboard. This conversion process is highly customizable, enabling users to modulate various instrument characteristics according to their preferences and facilitating a wide range of sound expressions. 
This flexibility aligns with the diverse potential applications of the tool, ranging from outreach activities—where aligning sound choices with personal preferences can enhance engagement—to education and research, where converging on standardized sounds can facilitate common understanding and analysis.
We note that in the current version of the code, customization can be performed via command-line arguments when launching the tool, making it compatible with speech-to-text software for blind and BVI users.


The default image-to-sound mapping in \texttt{herakoi} aligns with the most commonly used mappings in astronomical data sonification (\citealp{Zanella2022, Misdariis2022, Guiotto2024b}). 
It utilizes a physical analogy between wavelength and frequency to 
quantitatively
link specific colors in the target image to corresponding sound pitches --- mapping color to pitch, with red representing lower pitches and blue higher ones. The color scale is derived from the hue parameter in the HSB (Hue, Saturation, Brightness) model of the input image, truncated at 80\% of the total hue scale to avoid issues related to the cyclic nature of the HSB representation.
Since \texttt{herakoi} generates sound information as MIDI messages, the pitch values are discretized over the chosen tuning range. The brightness parameter is linearly mapped to the amplitude of the output sound, meaning that dimmer pixels correspond to quieter sounds and brighter pixels to louder ones, within a user-definable dynamic range for sound amplitude.

\begin{figure}
  \centering
\resizebox{1\hsize}{!}{\includegraphics[clip=true]{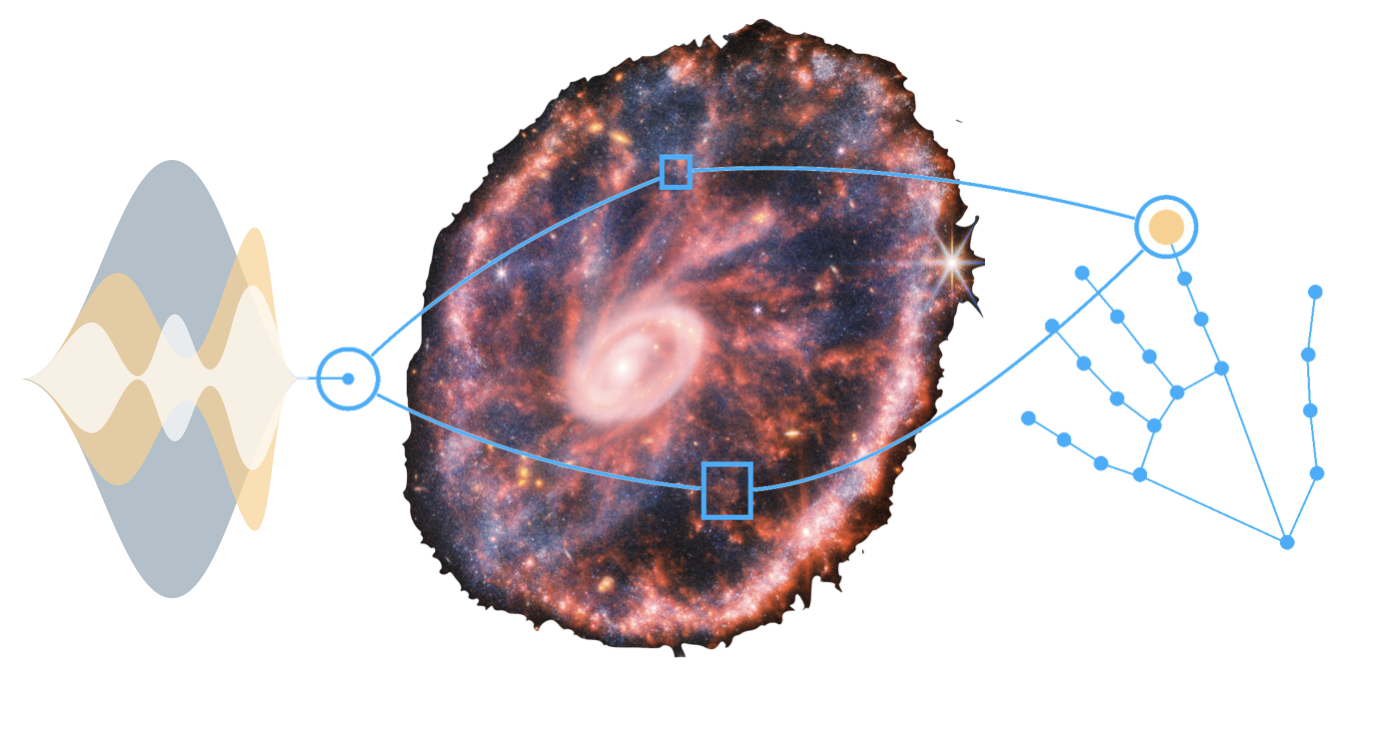} }
\caption{
\footnotesize
A visual representation of the \texttt{herakoi} sonification algorithm. 
The visualisation at the centre contains a composite image of the Cartwheel Galaxy observed with the James Webb Space Telescope (WebbTelescope.org: NASA, ESA, CSA, and STScI).}
\label{fig:sketch}
\end{figure}

\vspace{-0.3cm}
\section{\texttt{herakoi} in action}
\label{sec:discussion}

\texttt{herakoi} has been successfully used in varied contexts ––– from numerous public outreach events to academic workshops --- capturing the attention of both professional astronomers and the general public. Noteworthy cases include its integration in the ``Macchine del Tempo'' exhibition\footnote{\url{https://macchinedeltempo.inaf.it/index.php/homepage/}} by the Istituto Nazionale di Astrofisica at Palazzo delle Esposizioni in Roma, along with being showcased across multiple events in the context of the European Researchers' Night (e.g., Trieste Next\footnote{\url{https://www.triestenext.it/}}, Bright Night\footnote{\url{https://bright-night.it/}}) and at the yearly Astronomy Festival ``The Universe in all senses''\footnote{\texttt{\url{https://sites.google.com/inaf.it/festivaldiastronomia-eng/home}}} as part of the educational hands-on activity ``Let's play the stars!''\footnote{\texttt{\url{https://play.inaf.it/suoniamo-le-stelle/}}}.
Its ability to create an immersive, multi-sensory experience has proven to be a powerful way of communicating complex astronomical concepts, fostering greater engagement and understanding. 
The interactive nature of the tool has resonated well with users, including those who are BVI, helping to bridge the gap between visual-centric astronomy and inclusive, accessible science.
In addition to \texttt{herakoi}, a spin-off version called \texttt{edukoi}
\footnote{\url{https://github.com/herakoi/edukoi}}
was developed specifically for educational purposes (\citealp{Guiotto2023, Guiotto2024b}). 
\texttt{edukoi} builds on the core functionalities of \texttt{herakoi} but incorporates additional features aimed at enhancing the educational experience for schoolchildren, both sighted and BVI. 
\texttt{edukoi} has been tested in middle schools with around 150 students, ranging from 11 to 14 years of age, in a structured experiment designed to evaluate the tool’s effectiveness in color and shape recognition, as well as its impact on student engagement with science.
The tests, conducted in collaboration with the ``Scuola Media Don Milani'' in Italy, and reported by \cite{Guiotto2024b}, involved students interacting with basic geometric shapes and astronomical images using sound. The results demonstrated that \texttt{edukoi} was effective in helping students recognize colors\footnote{We note that the definition of ``color'' in this context refers to given specific properties of the sonified object. The lack of a visual support thus makes the method inherently independent of the actual chromatic aspects of the considered targets, in turn reducing potential biases and providing aids in the case of users with color vision deficiencies}, with an average accuracy of 86\% in initial trials and up to 93\% in follow-up tests. Shape recognition was more challenging, with performance varying but still significantly above random guessing. Despite technical setbacks, such as hardware issues at the testing site, the tool maintained high engagement levels, with students showing considerable enthusiasm and curiosity throughout the experiments. 
In addition, \texttt{edukoi}'s role in reshaping students’ attitudes toward science was assessed through pre- and post-intervention surveys. 
Qualitative feedback indicated that the hands-on, multisensory approach was both enjoyable and thought-provoking. 
Many students expressed increased empathy towards BVI individuals and a deeper understanding of the accessibility challenges in STEM fields.
\texttt{herakoi} and \texttt{edukoi} demonstrate the potential of sonification as a powerful tool not only for scientific research but also for education
\footnote{
We note that the quantitative mapping between image and sound properties would easily allow for research applications. However, to date, we have not investigated such a possibility in a systematic way, and we aim at pursuing this in future works.
}
.

\vspace{-0.5cm}

\section{Conclusions and future developments}

Data sonification is the process of converting data into sound, and it is taking on an increasingly pivotal role in science, research, education, and data accessibility (\citealp{Zanella2022, Harrison2022}). A relevance that has obtained a crucial, formal recognition in a special report\footnote{\url{https://www.unoosa.org/oosa/en/ourwork/space4personswithdisabilites/}} from the United Nations Office for Outer Space Affairs (UNOOSA).
This technique transcends traditional ways of sensing data, providing a unique sensory experience that is crucial for BVI people and offers an alternative perception for sighted individuals.
Sonification not only facilitates a novel way to process and experience data but also triggers new emotional and cognitive responses, thereby enhancing understanding and engagement with complex information.
The field of data sonification has the potential for significant growth, boosted by advancements in computer vision and AI technologies.
Leveraging these developments, we created \texttt{herakoi}, a software that stands at the forefront of this innovative field. \texttt{herakoi} is an AI-assisted tool designed to interpret and sonify visual elements, making data accessible in a multi-sensory manner. By performing real-time tracking of hand movements, \texttt{herakoi} allows the users to dynamically interact with images, translating visual properties of  touched pixels in images into auditory signals. 
\texttt{herakoi} has demonstrated significant potential, especially in the realm of astronomy, traditionally reliant on visual data. It has proven to be an effective educational tool in schools, helping young children grasp astronomical concepts. Also, \texttt{herakoi} has engaged diverse audiences during public outreach events, making it a precious resource for both sighted and BVI individuals. 

One of the most promising directions for improving \texttt{herakoi} lies in leveraging recent advancements in large language models (LLMs) and vision technologies.
We aim to combine \texttt{herakoi} with  large language and vision models, fine-tuned specifically for the astronomical domain (see e.g., \citealp{astrollama2024, pathfinder2024}), 
to enhance its capability of interpreting visual data and providing verbal insights.
This integration, coupled with voice synthesis, will allow \texttt{herakoi} to offer real-time insights and engage users—particularly BVI individuals—in interactive dialogue about image content.
By enabling verbal descriptions and answering user queries, \texttt{herakoi} will bridge gaps in data accessibility, providing a richer, more autonomous experience.




\begin{acknowledgements}
LDM has been supported by the French government, through the UCA\textsuperscript{J.E.D.I.} Investments in the Future project managed by the National Research Agency (ANR) with the reference number ANR-15-IDEX-01.
\end{acknowledgements}

\bibliographystyle{aa}
\bibliography{bibliography}

\end{document}